\documentclass[11pt,a4paper,twocolumn]{article}

\usepackage{natbib}

\usepackage{bm}
\usepackage{amssymb}
\usepackage{xcolor}

\usepackage{graphicx}	

\def\Red{}
\def\redc{}

\def\a{\alpha}

\def\d{\delta}
\def\D{\Delta}

\def\l{\lambda}

\def\M{{\cal M}}
\def\rr{\varrho}
\def\g{\gamma}
\def\G{\Gamma} 

\def\O{\Omega}
\def\th{\theta}

\def\s{\sigma}

\def\ra{\rightarrow}

\begin{document}


\title{Scaling laws in the stellar mass distribution and the transition to homogeneity}

\author{Jos\'e Gaite
\\
\small\em
Physics Dept.,
ETSIAE, Univ.\ Polit\'ecnica de Madrid, E-28040 Madrid, Spain
}

\date{\today}
\maketitle

\begin{abstract}
We present a new statistical analysis of 
the large-scale stellar mass distribution 
in the Sloan Digital Sky Survey (data release 7). 
A set of volume-limited samples shows that the stellar mass of galaxies 
is concentrated in a range of galaxy luminosities that is very different 
from the range selected by 
the usual analysis of galaxy positions. 
Nevertheless, the
two-point correlation function is a power-law with the usual exponent $\g=1.71$--$1.82$, which varies with luminosity. 
The mass concentration property allows us 
to make a meaningful analysis of the angular distribution of the full flux-limited sample. 
With this analysis, after suppressing the shot noise, 
we extend further the scaling range and thus 
obtain $\g=1.83$ and a clustering length \redc{$r_0= 5.8$--$7.0\,h^{-1}$Mpc.} 
Fractional statistical moments of the coarse-grained stellar mass density  
exhibit multifractal scaling. 
Our results support a multifractal model with a transition to 
homogeneity at about $10\,h^{-1}$Mpc. 
\end{abstract}



\section{Introduction}
\label{intro}

It is claimed that we are entering the era of precision cosmology, in which the 
fundamental parameters of the universe are known within a few percent precision and only 
remains to progressively refine them \citep{PrecCosmo,PDG}. In addition to 
the global parameters of the FLRW model,
we have the parameters that determine 
the primordial density perturbations and hence 
the formation of large scale structure. 
A particularly interesting  parameter is the amplitude of the primordial density perturbations, usually measured 
in terms of the present linear-theory mass dispersion on a scale of 8 $h^{-1}$Mpc,
named $\s_8$.
The scale of 8 $h^{-1}$Mpc comes from Peebles' observation that galaxy counts on this 
scale have a rms fluctuation 
approximately equal to one \citep{Peebles}. 
In general, the mass dispersion over the length $R$,
$\s(R)$, defines a scale such that $\s(R)=1$. 
This scale  
separates the linear and quasi-homogeneous regime of the evolution of density fluctuations from the nonlinear regime of strong clustering. 

A similar scale is the 
distance $r_0$ such that $\xi(r_0)=1$,
where $\xi(r)$ is the reduced galaxy-galaxy correlation function. 
Actually, this function is well approximated by the power law
\begin{equation}
\xi(r)=
\left(\frac{r_0}{r}\right)^{\gamma},
\label{xi}
\end{equation}
for distances not much larger than $r_0$ \citep{Peebles}.
The scale such that \Red{the rms dispersion of galaxy counts is equal to one}
is related to $r_0$ 
by a $\g$-dependent factor \citep{Peebles}. 
Historically, the power-law form of $\xi(r)$
was 
deduced from the {\em angular} positions of galaxies, yielding $\g=1.8$
and $r_0 = 4.7\,h^{-1}$Mpc \citep{Tot-Ki,Groth}. 
Nowadays, good galaxy redshift surveys are available, but 
the angular positions are still useful. 
In particular, the angular correlation function $w(\th)$ is used, in combination with other 
data, to determine precision values of $\s_8$ \citep{PDG,DES}.

The amplitude of the primordial density perturbations 
is not theoretically constrained and 
determines the present scale of transition to homogeneity, which 
can have any value, once given the global cosmological parameters.  
The transition to homogeneity has been the subject of numerous studies, 
some of them motivated by the bold proposal that the scale of homogeneity is too large 
to be accessible or even that no such transition exists 
\citep{Mandel,Pietronero,Cole-Pietro}. This proposal, namely, that the mass distribution is
fully scale invariant, at all scales in a Newtonian cosmology, has been called the 
``fractal universe'' \citep{Peebles}. 
Actually, Eq.~(\ref{xi}) corresponds to a fractal distribution on scales $r \ll r_0$, 
for any magnitude of $r_0$. 

The scale and specific form of transition to homogeneity determine the size of the largest structures, although these structures can have a length much larger than $r_0$ \citep{I0}. 
An examination of the literature shows values of \Red{the scale of transition to homogeneity} 
that go from the standard value of 5 $h^{-1}$Mpc \citep{Peebles} to more than 20 $h^{-1}$Mpc 
\citep{SyL-Bar,Sar-Yad,Verev,CSF}. 
The range that these values span is quite long. 
In contrast, the current precision values of $\s_8$  
are assigned a relative error of about one percent \citep{PDG}. This precision 
is surprisingly good, in comparison, given that $\s_8$ determines 
the scale of transition to homogeneity. 
 
Most of the literature about the large-scale structure of the universe based on 
the statistical analysis of galaxy catalogs is limited to the 
distribution of the galaxy {\em positions}. 
We make here, following up on \citet{I-SDSS}, 
a new statistical analysis of the large-scale structure, 
adding an important ingredient: the {\em stellar masses} of galaxies; that is to say, 
we study 
the large-scale distribution of stellar mass.
The distribution of stellar mass in the SDSS has already been studied by \citet{Li-W}, 
employing the projected correlation function $w(r_\mathrm{p})$, 
on scales $r_\mathrm{p} < 30\,h^{-1}$Mpc 
[$r_\mathrm{p}$ is the separation perpendicular to the line of sight \citep{DaPe}].
\citeauthor{Li-W} find that 
$w(r_\mathrm{p})$ 
is very well represented, over the range 
$10\,h^{-1}$kpc $<r_\mathrm{p}<$ $10\,h^{-1}$Mpc, 
by a power law that corresponds to Eq.~(\ref{xi}) with 
$\g=1.84$ and $r_0 = 6.1\,h^{-1}$Mpc. 
These results basically agree with the analysis of 
the SDSS galaxy positions by \citet{Zehavi}, 
but \citeauthor{Li-W} deem the scaling of the stellar mass to be better.  
At any rate, the stellar mass and galaxy number distributions can be shown to be very different, 
thus warranting further statistical analysis.	

Besides, we must take into account that 
the ``projection'' that produces 
$w(r_\mathrm{p})$ is performed on the correlation function $\xi(r)$ and is not associated to 
an actual projected distribution but  
to a set of local orthogonal projections on the tangent planes to the surface of the unit sphere (a rather unwieldy construction from the mathematical standpoint).
Arguably, the direct study of the angular projection of the stellar mass distribution 
is preferable. 
On the other hand, there are statistical measures not based on the two-point correlation
function.
Our study combines the analysis of 
a set of volume-limited samples
with an angular analysis and, 
moreover, involves various statistical measures.

The importance of galaxy masses in the study of 
galaxy clustering was recognized years ago \citep{Pietronero}. 
Furthermore, \citet{Pietronero} argued that a full {\em multifractal} analysis 
of galaxy clustering is necessary. 
A multifractal model was also proposed by \citet{Jones-88}, 
without considering the galaxy masses.
Pietronero and collaborators initiated the multifractal analysis with 
masses of galaxies derived from the observed luminosities
by assuming a simple mass-luminosity relation 
\citep{Cole-Pietro,Sylos-PR}.
Meanwhile, 
the large-scale structure has been known to consist not just of clusters but of 
a {\em web structure} \citep{Ge-Hu,Kof-Pog-Sh-M}. This structure can be 
described as a multifractal \citep{I-SDSS,galax2}.
Multifractality has become essential to the scaling analysis 
of 
the distribution of galaxies, 
although normally without considering the galaxy masses \citep{Borgani,Jones-RMP}.

As we shall see, most of a multifractal structure is preserved in 
an angular projection. 
\citet{Pietronero} already
considered some basic properties of the angular projection 
of a fractal galaxy distribution. This question 
was further studied by \citet{Cole-Pietro} and \citet{Durrer}, assuming 
a monofractal model.


Our galaxy data come from the Sloan Digital Sky Survey, 
data release 7 (SDSS DR7) \citep{SDSS-DR7}, 
as provided by the New York University Value-Added Galaxy Catalog (NYU-VAGC) \citep{VAGC}.
The NYU-VAGC contains 
the stellar masses of the galaxies, calculated with the method described by \citet{Blan-Ro} 
(which gives similar results to the method of \citealt{Kauff}).
In fact, we use the same data as \citet{Li-W}, which facilitates the comparison of the 
respective results. 
Moreover, 
the SDSS DR7 is well studied and, in particular, there is a careful analysis
of its galaxy-galaxy angular correlation function \citep{Wang-etal}. 
We have already employed the same data to make a multifractal analysis 
of the stellar mass distribution based on the convergence of multifractal spectra 
for a shrinking scale
\citep{I-SDSS}. 
The present treatment of scaling laws is much more elaborate, 
allowing us to compare with the preceding studies of scaling in the SDSS and to
study the transition to homogeneity. 

Our plan is as follows.
We begin in Sect.~\ref{z} with a short discussion of scaling, \Red{fractality} and homogeneity 
in galaxy surveys and of the role played by the scale-dependent mass variance. 
In Sect.~\ref{VL}, 
we apply these concepts to the study of a set of volume-limited samples from the SDSS DR7 
and to the analysis of scaling of the respective mass variances. 
Multifractality is proved by the analysis of fractional statistical moments in 
Sect.~\ref{frac}. 
We next consider the theory of projection of fractal distributions 
in Sect.~\ref{Fp}, and 
we connect this theory with the standard theory of the angular 
two-point correlation function of galaxies 
(Sect.~\ref{angular}). 
After this theoretical study, we undertake the analysis of the 
angular projection of the SDSS DR7 in Sect.~\ref{angSDSS}, where 
we deal with the increased shot noise (Sect.~\ref{shot}) and 
we obtain $r_0$ from the variance of the coarse-grained angular density (Sect.~\ref{r0}). 
We end with 
a general discussion (Sect.~\ref{discuss}).

\section{Scaling, fractality and homogeneity in galaxy surveys}
\label{z}

It is pertinent to begin with some general considerations about scaling and homogeneity in 
galaxy surveys, and, in particular, about how to characterize fractality and how to 
determine the scale of transition to homogeneity. 
Let us assume that we can obtain the three-dimensional stellar mass distribution, 
which requires 
a redshift survey with galaxy stellar masses and 
the construction of volume limited samples.

Taking the two-point correlation function as the basic statistic, the fundamental scaling law 
is the power-law reduced correlation function in Eq.~(\ref{xi}). This law is 
supposed to hold when $\xi(r)$ is not small, that is to say, 
for $r$ not much larger than $r_0$ (but it could also be valid for $r\gg r_0$). 
The coarse-grained mass fluctuation $\d M_R/M_R$,
namely, the scale dependent rms fluctuation of the stellar 
mass \Red{in a cell of linear dimension $R$}, derives from 
the reduced two-point correlation function of the mass distribution. 
If this function 
follows the power law (\ref{xi}), the mass variance follows a power law with the same exponent, namely, 
\begin{equation} 
\left(\frac{\d M_R}{M_R}\right)^{2} = C(\g)\left(\frac{r_0}{R}\right)^{\gamma},
\label{dMR}
\end{equation} 
where the form of $C(\g)$ depends on the shape of the cell 
\citep{Peebles}.  

Regardless of any scaling property, $\d M_R/M_R$ must decrease 
with $R$ and 
the scale \Red{of homogeneity} can be defined just by the condition $\d M_R/M_R=1$.   
However, this is too stringent a criterion for homogeneity. 
Indeed, 
let us notice that a positive random variable with a rms dispersion 
equal to its mean value cannot be approximately Gaussian; 
for example, let us consider the lognormal distribution \citep[see][p.\ 5]{lognormal}.
A suitable criterion for homogeneity is the mass variance $(\d M_R/M_R)^2=0.1$, 
as assumed by \citet{I-SDSS}; indeed, a
lognormal distribution function with this variance is hardly skewed 
\citep[p.\ 5]{lognormal}.
Whether or not the scale defined by $(\d M_R/M_R)^2=0.1$ is close 
to the one defined by $\d M_R/M_R=1$ 
depends on 
how sharp the transition to homogeneity is. To analyze this question, 
we naturally choose the power law form in Eq.~(\ref{dMR}).

For a spherical  cell of radius $R$, $C(\g)$ in Eq.~(\ref{dMR}) is given by 
\citet{Peebles}. 
The scale $R$ such that $\d M_R/M_R$ equals a given number is a definite function of 
$\g$.
For $\d M_R/M_R=1$, $R$ is only a little larger than $r_0$, whereas, 
for $(\d M_R/M_R)^2=0.1$, 
it is between $4.7r_0$ and $12r_0$ (the larger, the smaller $\g$ is). 
In conclusion, the scale such that $\d M_R/M_R=1$ is roughly equivalent to $r_0$ 
but the scale where the probability distribution is approximately Gaussian 
can be several times larger.  
This observation
reveals that the values $r_0=5.4\,h^{-1}$Mpc and $\g=1.8$ 
are compatible with the presence of relatively large structures, for example, 
cosmic voids of size $\simeq 30\,h^{-1}$Mpc \citep{Peebles}. Actually, even larger structures 
can be observed, provided that $\xi(r)$ does not fall too rapidly 
for $r \gg r_0$ \citep{I0}.

After clarifying our concept of homogeneity, 
let us now recall several aspects of scale invariance and fractal geometry.
First of all, let us remark that the fractal geometry of a mass distribution 
consists in some sort of scale invariance in the strongly inhomogeneous or 
strong clustering regime. 
Its most general form 
is called multifractality and is 
expressed in terms of the multifractal spectrum $f(\a)$ or, alternatively, of 
the {\em R\'enyi dimension} spectrum $D_q$ \citep{Harte,Falcon}. 
The multifractal spectrum $f(\a)$ is a basic function 
that represents how mass concentrates, in terms of 
the number of mass concentrations of a given ``strength'', measured by the {\em local 
dimension} $\a$, while the R\'enyi dimensions provide a sort of averaged information. 
The functions $f(\a)$ and $D_q$ are related by a Legendre transform in terms of 
the variables $\a$ and $q$.
The R\'enyi dimension $D_q$ expresses the scaling of the 
statistical $q$-moment of the coarse-grained mass distribution 
and is, in general, a non-increasing function of $q$. 
The 
second order moment 
has an intuitive meaning and
the corresponding R\'enyi dimension is $D_2=3-\g$.  
General $q$-moments and their scaling relations are studied in Sect.~\ref{frac}.

It is to be remarked that the primary concept of fractal dimension and its various definitions 
are concerned with the small scale behavior of sets or 
mass distributions, and fractality is roughly equivalent to {\em asymptotic} scaling 
in the limit of vanishing scale, as described in standard fractal geometry textbooks 
\citep{Harte,Falcon}. 
In consequence, multifractal measures
are defined for {\em any} mass distribution, regardless of its behavior on large scales. 
An arbitrary mass distribution is not expected to be such 
that it becomes homogeneous on large scales, tending to 
a uniformly homogeneous state, as occurs in cosmology because of the cosmological principle. 
This is why raw statistical $q$-moments $\langle M^q\rangle$ 
are employed in fractal geometry, instead of {\em central} moments, 
which require a globally defined mean density. 

To make the cosmological principle compatible with fractal models of strong clustering, 
\citet{Mandel} proposed the {\em conditional cosmological principle}, 
in which ``every possible observer'' is replaced with 
``every observer located at a material point''. 
Thus,  
the natural measure for the fractal analysis of 
strong clustering is the conditional density, 
\Red{namely, the average density at a distance $r$ from an occupied point}
\citep{Mandel,Cole-Pietro,Sylos-PR}. 
It can be expressed as
\begin{equation} 
\G(r)=\frac{\langle \rr(\bm{r})\,\rr(\bm{0})\rangle}{\langle \rr\rangle}
= \langle \rr\rangle \left(\xi(r)+1\right).
\label{Gcond}
\end{equation}
But it is conceptually independent of the mean density
$\langle \rr\rangle$, is spite of the fact that 
it appears in these expressions.

Motivated by 
these ideas, 
some authors look for scaling of the conditional density $\G(r)$ of the galaxy distribution   
(or of its integral in the ball of radius $r$). 
When calculated with this method, the scale of homogeneity is often large 
\citep{SyL-Bar,Verev}.
The conditional density 
is not used by traditional cosmologists, who are accustomed to the function $\xi(r)$.  
If $\langle \rr\rangle \neq 0$, then the conditional density $\G(r)$ is conceptually useful 
but can be expressed in terms of $\xi(r)$, by Eq.~(\ref{Gcond}), and it scales as 
$\xi(r)+1$.
Naturally, the condition for {\em fractal scaling} is 
$\xi \gg 1$, which makes \Red{$\xi$ and $\xi + 1$} equivalent.
However, the condition is never fulfilled in a sufficiently long range of $r$.
Therefore, the values of $r_0$ and $\g$ obtained from power-law fits of either 
function differ, namely, $r_0$ is larger when calculated from $\xi+1$. 
Let us illustrate this point with 
an example, employing a sample from \citet{I-SDSS}.

\begin{figure}
\centering{
\includegraphics[width=8cm]{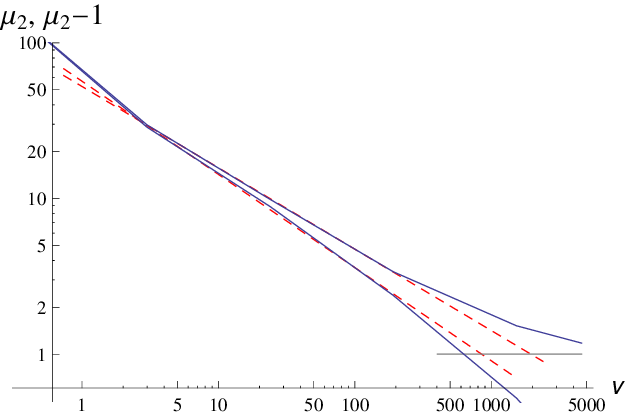}
}
\caption{Fits to scaling
for sample VLS1: moments $\mu_2(v)$ and $\mu_2(v)-1$ 
($v$ in Mpc$^3/h^3$), with fits to the scaling $(v/v_0)^{-\g/3}$, 
\Red{represented as dashed lines.} 
The crossing \Red{of each dashed line} with the horizontal gray line gives $v_0$.
}
\label{fits}
\end{figure}

\citet{I-SDSS} uses a method of multifractal analysis based on 
coarse-graining cells that are adapted to the SDSS sample geometry. 
These cells are not spherical nor have a simple 
shape, but have equal volume, which serves to measure their size. 
Let this volume be $v$. 
The second moment of the coarse-grained density is 
\begin{equation} 
\mu_2(v) = \frac{\langle \rr_v^2\rangle}{\langle \rr_v \rangle^2}\,,
\label{mu2}
\end{equation} 
and the density variance (mass variance) is
\begin{equation}
\mu_2(v)-1 = \frac{\langle {\d\rr_v}^2\rangle}{\langle \rr_v \rangle^2}
= \frac{\langle {\d M_v}^2\rangle}{\langle M_v \rangle^2}\,.
\end{equation} 
The scaling law (\ref{xi}) is equivalent to the scaling law for the mass variance
(\ref{dMR}). 
Likewise, a scaling law for 
$\xi(r)+1$ is equivalent to a scaling law for $\mu_2(v)$. 
An example of both the scaling laws for $\mu_2(v)$ and $\mu_2(v)-1$ is displayed in 
Fig.~\ref{fits}, 
which we now explain.

Fig.~\ref{fits} refers to 
a volume limited sample \Red{of 1765 galaxies in the} 
redshift range $(0.003,0.013)$, defined by \citet{I-SDSS} and called VLS1. 
This sample was  
adequate to compute a quite complete multifractal spectrum and is now useful to 
test the two options for the scaling law that yields $D_2$. 
The solid blue lines in Fig.~\ref{fits} are the graphs of $\mu_2$ and $\mu_2 -1$. 
The fit to the power law 
\begin{equation} 
\mu_2 = (v/v_0)^{-\g/3}
\end{equation} 
in the interval $v\in [3, 190]$ Mpc$^3$/$h^3$ 
yields $v_0=(1960 \pm 180)$ Mpc$^3$/$h^3$ and $\g=1.57\pm 0.01$, 
that is to say, $D_2=3-\g=1.43\pm 0.01$ \Red{(the power-law fit is a least-squares 
linear fit of $\log\mu_2$ versus $\log v$)}.
An analogous fit to the power law 
\begin{equation} 
\mu_2 -1 = (v/v_0)^{-\g/3}
\label{plaw}
\end{equation} 
yields $v_0=(850 \pm 300)$ Mpc$^3$/$h^3$ and $D_2=1.20\pm 0.06$
($\g=1.80\pm 0.06$). Both fits are represented by dashed red lines in Fig.~\ref{fits}.
The latter fit, with $\g=1.80$ and $v_0^{1/3}=9.5$ Mpc/$h$, basically agrees 
with the canonical values of $\g$ and $r_0$ (from the reduced two-point 
correlation function of galaxy {\em positions}). 
However, both fits are questionable for calculating $D_2$, because the fractal regime 
demands $\mu_2-1 \approx \mu_2 \gg 1$, but
the \Red{fitted range} extends beyond the small scales where this condition holds.
Therefore, some intermediate values of $v_0$ and $\g$ should be more appropriate.  
These values are 
uncertain \Red{because we have too small a range of scales
with both $\mu_2 \gg 1$ and negligible discreteness effects.} 
These effects make $D_2$ shrink 
towards $D_2=0$, 
as the steeper left ends of the graphs reflect. 

The uncertainties of 20\% in $D_2$ and 30\% in $v_0^{1/3}$ due to 
the limited scaling range
can be compared 
with the uncertainties 
in galaxy positions and masses. 
\citet{I-SDSS} has analyzed how these uncertainties affect the multifractal spectrum of 
VLS1, with the result that only the right-hand side part of the 
spectrum is affected \citep[left-hand side of Fig.~4]{I-SDSS}. 
This implies that only the (R\'enyi) dimensions $D_q$ with $q<0$ are affected. 
Therefore, the evaluation of $D_2$ should not be affected. 
To confirm it, we \redc{calculate} now the uncertainty 
of $\mu_2(v)$ due to the uncertainty of galaxy positions and masses. 
To do it, we employ the same \redc{procedure} as \citet{I-SDSS}, namely, we generate ten 
variant samples, with values of redshift and stellar mass given by, respectively,  
Gaussian and lognormal distributions with variances in accord with the literature.  
The variant samples yield ten values of $\mu_2$ for each value of $v$, and we compute 
for each $v$ the standard deviation \redc{of} those ten values. The result is that 
the relative error of $\mu_2$ is always smaller than 4.2\% 
(the corresponding error bars in Fig.~\ref{fits} would hardly be visible).

As regards errors in the linear fits, their magnitude 
grows as one tries to fit more points, and the natural 
criterion to stop is to minimize the rms error per degree of freedom 
(the number of degrees of freedom is the number of points minus two).
In most of the fits here, 
these errors are quite small in comparison with errors from other sources.
The relative small uncertainty due to statistical errors in galaxy positions and masses combined with the smallness of errors in the linear fits shows that 
the major uncertainty in the fractal analysis is due to 
discreteness effects, that is to say, to limitations imposed by 
a small fractal scaling range. 

\citet{I-SDSS} observes that the low-redshift VLS1 
obtains a multifractal spectrum that is more complete than the ones obtained from 
higher-redshift volume-limited samples. 
We find that higher-$z$ samples are also less useful to directly calculate $D_2$ as above. 
The \Red{higher $z$ is in a} volume-limited sample the 
more luminous the galaxies in it and the smaller their number density 
(as we shall discuss when we construct a set of volume-limited samples in Sect.~\ref{VL}). 
A smaller number density implies that discreteness effects 
take over growing ranges of the smaller scales, reducing  
the interval of scales where the fractal regime can be measured, 
that is to say, the interval where $\mu_2-1 \approx \mu_2 \gg 1$. 

However, 
we may consider the scaling of just $\mu_2(v)-1$, in the longest possible interval,   
without demanding that $\mu_2(v)-1 > 1$.  
For example, one
can dismiss the upper graph in Fig.~\ref{fits}, corresponding to $\mu_2$, 
and think of extending the scaling of $\mu_2-1$ below the line $\mu_2-1=1$. 
Likewise,  
Eq.~(\ref{xi}) usually extends to $\xi<1$ (weak correlations). 
Besides, 
we remark that the frequently employed projected correlation function 
$w(r_\mathrm{p})$ (e.g., by \citealt{Zehavi} or \citealt{Li-W})
is a {\em dimensional} function that renders obscure
the strength of correlations.  
%
The range with $\xi<1$ is the appropriate one for the molecular correlations 
in statistical physics, 
and scale invariance takes place in {\em critical phenomena}.
In this regard, we can expect that Eq.~(\ref{xi}) holds from the small scales 
where $\xi\gg 1$ to the large scales where $\xi\ll 1$, encompassing the fractal and {\em critical-fluid} regimes 
\citep{I0}.

At any rate, what we conclude from the analysis of VLS1 is that 
the statistical analysis of the stellar mass distribution in the \Red{SDSS-DR7} by means of 
volume-limited samples is plagued with errors, especially, discreteness errors. 
An alternative to it is 
the study of the projected angular distribution,  
which has two advantages. 
On the one hand, we avoid the influence of redshift space distortions on the 
statistics, which is considerable \citep{Zehavi}.  
On the other hand, we have all points in the angular region, whereas  
volume-limited samples 
span the same angular region but 
occupy very different three-dimensional regions (which are awkwardly shaped, in addition). 
%
We see next the information obtained from a systematic study of 
volume-limited samples and leave the 
study of the angular projection to Sect.~\ref{Fp}.

\section{Volume-limited samples}
\label{VL}

To be systematic in the study of the three-dimensional mass distribution, 
we construct volume-limited samples
in successive intervals of absolute magnitude; for example, 
in the SDSS, one can take intervals in which it changes by one unit. 
We follow the procedure and conventions of \citet{Zehavi}   
(the absolute magnitude is denoted by $M_\mathrm{r}$, not to be confused with our notation of 
$M$ as a mass).
 
We employ the \Red{same} SDSS-DR7 data from the NYU-VAGC 
as in 
\citet{I-SDSS}, and we also \Red{select}
the apparent magnitude range $12.5 < m_\mathrm{r} < 17.77$. The angular ranges 
are defined in terms of equal-area coordinates 
$sl=\sin \l$ and $f=\eta+0.567$ radians, where the angles $\l$ and $\eta$ 
constitute the SDSS coordinate system. 
\Red{The available ranges of our coordinates are} $-0.760<sl<0.793$ and $-0.0227<f<1.200$ 
\Red{(radians), providing} 
a total solid angle $\Omega=1.554 \cdot 1.222=1.899$ steradians. 
We further restrict the sample to $z<0.34$, obtaining 
$529\hspace{1pt}655$ galaxies.





 \begin{table}[h]
 \begin{center}
 \scriptsize
    	\begin{tabular*}{\linewidth}{@{}c@{~}c@{~}r@{~}c@{~}c@{~}c@{}}
 	\hline
	$M_\mathrm{r}$ & $z$ & $N$ & 
$n$ & dens.\ & mass \\
	\hline\hline
	$(-14,-15)$ & 0.0011--0.0073 & 554 & 
0.08550 & 0.009 & $(2.7~10^4,1.4~10^8)$\\
	\hline
	$(-15,-16)$ & 0.0017--0.0115 & 885 & 
0.03472 & 0.010 & $(4.1~10^5,4.0~10^8)$\\
	\hline
	$(-16,-17)$ & 0.0026--0.0181 & 1708 & 
0.01717 & 0.012 & $(9.4~10^6,9.2~10^8)$\\		
	\hline
	$(-17,-18)$ & 0.0042--0.0284 & 6625 & 
0.01728 & 0.039 & $(1.7~10^6,2.8~10^9)$\\
	\hline
	$(-18,-19)$ & 0.0066--0.0444 & 17416 & 
0.01202 & 0.089 & $(1.2~10^7,6.1~10^9)$\\
	\hline
	$(-19,-20)$ & 0.0105--0.0689 & 46204 & 
0.00870 & 0.203 & $(2.1~10^7,2.3~10^{10})$\\
	\hline
	$(-20,-21)$ & 0.0165--0.1058 & 97847 & 
0.00522 & 0.333 & $(1.7~10^6,6.6~10^{10})$\\
	\hline
	$(-21,-22)$ & 0.0260--0.1613 & 101350 & 
0.00159 & 0.247 & $(1.2~10^7,1.7~10^{11})$\\
	\hline
	$(-22,-23)$ & 0.0406--0.2444 & 30680 & 
0.00015 & 0.056 & $(7.5~10^8,4.4~10^{11})$\\
	\hline
	$(-23,-24)$ & 0.0632--0.3597 & 1460 & 
$2.4~10^{-6}$ & 0.002 & $(7.7~10^9,7.9~10^{11})$\\
	\hline
        \end{tabular*}
 \caption{Characteristics of the volume-limited samples: absolute magnitude, redshift, 
number of galaxies, 
number density ($h^{3}$ Mpc$^{-3}$), fraction of the total mass density, and 
galaxy mass ($M_\odot$).} 
  \label{VLStable}
 \end{center}
 \end{table}

The main characteristics of our volume-limited samples are reported in Table~\ref{VLStable}. 
Naturally, the redshift intervals are almost coincident with the ones of 
\citet{Zehavi},  
but the the number of galaxies in each sample is larger, 
because the angular region is larger.
%
The number of galaxies grows with absolute luminosity, up to $M_\mathrm{r} \in (-21,-22)$, 
with $N=101\hspace{1pt}350$. This growth might suggest that discreteness effects are reduced 
up to this point, 
but 
the number density shrinks and 
the overall effect is that discreteness progressively 
hinders the detection of strong clustering, as noticed in Sect.~\ref{z}. 
For instance, the sample with 
$M_\mathrm{r} \in (-21,-22)$ and number density $0.00159$ has a volume per galaxy of 
$629$ Mpc$^{3}\!/h^{3}$, that is to say, the equivalent length is 
$8.6$ Mpc$/h$, as large as the expected value of $r_0$. 

As regards number density, the best sample is the first one, 
with $M_\mathrm{r} \in (-14,-15)$ and number density $0.08550$ (this sample is somewhat 
like the VLS1 studied in Sect.~\ref{z}). However, we are 
studying here the distribution of mass rather than the distribution of individual 
galaxies and the sample represents 
a fraction of the total mass density that only amounts to $0.009$. Thus, 
the clustering properties of the full mass distribution are hardly influenced by the 
galaxies with $M_\mathrm{r} \in (-14,-17)$, in spite of their abundance.

The last variable in Table~\ref{VLStable} is the galaxy mass, and we can observe the 
evident correlation between galaxy mass and absolute luminosity. This correlation is 
significant in the multifractal analysis, because this analysis distinguishes the 
strength of mass concentrations, measured by the local dimension $\a$. 
To be precise,  
the concentration strength is measured in coarse multifractal analysis 
by the logarithm of the mass in each cell, 
namely, $\a = -\log m/\ln l$, 
(\citealt{Falcon}, applied to cosmology by \citealt{I-ApJ}). 
However, 
galaxies are not equal-size mass concentrations and, actually, the 
galaxy sizes are not even part of the data. 
Nevertheless, it is to be expected that the stellar mass in the more luminous samples is  
more concentrated. Therefore, we also expect that the set of values of $\a$ 
and hence the dimension $D_q$ of each sample are smaller the more 
luminous the galaxies in it.

\subsection{Scaling of cell mass variances}
\label{cell_var}

Here we examine if cell mass variances 
are power-law functions of the cell volume $v$, according to Eq.~(\ref{plaw}) and 
employing the coarse-graining method of \citet{I-SDSS} as in Sect.~\ref{z}. 
We focus on the three galaxy samples in Table~\ref{VLStable} in the interval 
$M_\mathrm{r} \in (-19,-22)$, 
whose contribution to the total mass density is dominant (more than 78\%).   
But we also examine, to be thorough, the two 
adjacent samples, to include 
most of the total mass density as well as most galaxies in the set of samples.

 \begin{table}[h]
 \begin{center}
 \footnotesize
    	\begin{tabular}{ccc}
 	\hline
	$M_\mathrm{r}$ & $\g$ & $v_0^{1/3}$ \\
	\hline\hline
	$(-18,-19)$ & $1.52 \pm 0.07$ & $13.5 \pm 1.6$ \\
	\hline
	$(-19,-20)$ & $1.71 \pm 0.04$ & $12.4 \pm 0.8$ \\
	\hline
	$(-20,-21)$ & $1.616 \pm 0.013$ & $14.2 \pm 0.3$  \\
	\hline
	$(-21,-22)$ & $1.818 \pm 0.007$ & $16.9 \pm 0.2$  \\
	\hline
	$(-22,-23)$ & $1.93 \pm 0.02$ & $29.4 \pm 0.6$ \\
	\hline
        \end{tabular}
 \caption{Results for scaling in the five volume-limited samples with higher mass 
density ($v_0^{1/3}$ in Mpc$/h$).} 
  \label{VLSresults}
 \end{center}
 \end{table}

\begin{figure}
\centering{
\includegraphics[width=8cm]{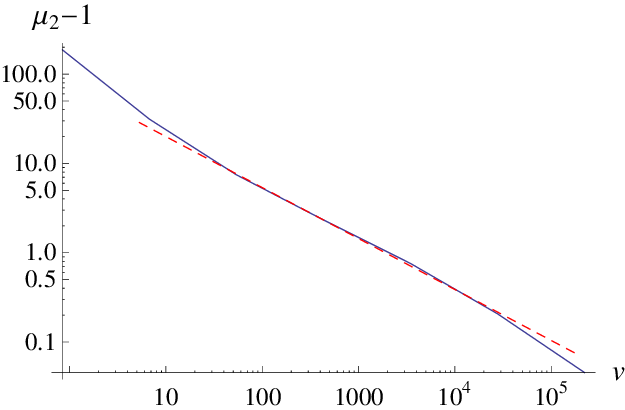}
\vskip 0.6cm
\includegraphics[width=8cm]{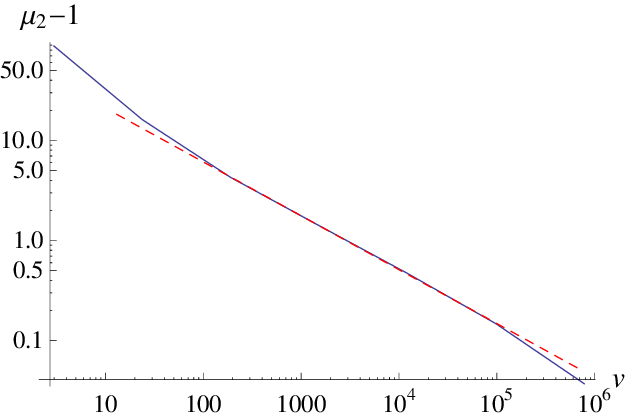}
\vskip 0.6cm
\includegraphics[width=8cm]{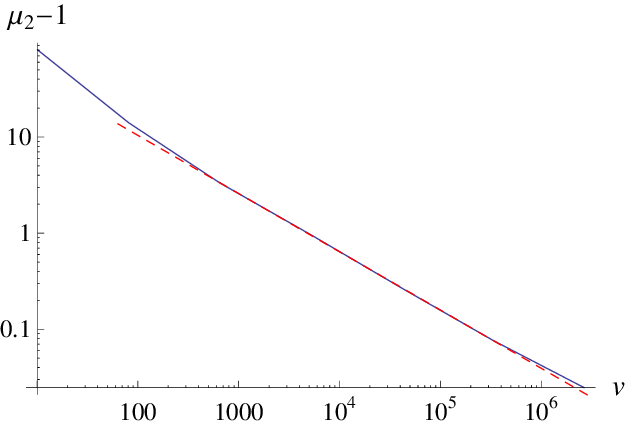}
}
\caption{
Variance $\mu_2(v)-1$ 
of the stellar mass distribution ($v$ in Mpc$^3/h^3$), 
for three volume-limited samples: $M_\mathrm{r} \in (-19,-20)$ (upper plot), 
$M_\mathrm{r} \in (-20,-21)$ (middle plot),
and $M_\mathrm{r} \in (-21,-22)$ (lower plot).
The \Red{linear} fits 
yield the values 
of $\g$ and $v_0^{1/3}$ in Table~\ref{VLSresults}. 
}
\label{ang-corr_VL}
\end{figure}

The results of the calculation and plotting of 
$\mu_2(v)-1$ appear in Table~\ref{VLSresults} and Fig.~\ref{ang-corr_VL}.
%
To wit, in Fig.~\ref{ang-corr_VL} are the log-log plots of $\mu_2(v)-1$ 
for the three samples with $M_\mathrm{r} \in (-19,-22)$, and in Table~\ref{VLSresults} 
are the numerical results of the log-log linear fits corresponding to 
the five volume-limited samples of Table~\ref{VLStable} with higher mass density. 
We obtain a nontrivial scaling for intermediate values of $v$, 
which crosses over, for small $v$, to the trivial scaling corresponding to isolated points ($D_2=0$). 
It is remarkable that the scalings displayed in Fig.~\ref{ang-corr_VL} hold 
in considerable ranges that go across the point of unit variance,  
extending from the moderately strong clustering regime to the quasi-homogeneous regime. 

Regarding the values of the scaling exponent $\g$ in Table~\ref{VLSresults}, 
we can observe a definite growing trend, except in the sample with 
$M_\mathrm{r} \in (-20,-21)$ (an anomaly in this range is also noted by 
\citealt{Zehavi}). 
The length scale $v_0^{1/3}$, which we can take as a homogeneity scale, also grows 
with luminosity. 
The growing trends of both $\g$ and the homogeneity 
scale are also obtained in the analysis of galaxy positions \citep[table 1]{Zehavi}. 

In the (moderately) strong clustering regime, the scaling of the mass variances 
denote fractality, 
with $D_2= 3 -\g \in (1,1.5)$, the lower values 
for the larger luminosities. 
This decreasing trend of fractal dimension as a function of luminosity is 
expected, as explained at the end of Sect.~\ref{VL}. 
Let us notice that the galaxies with 
$M_\mathrm{r} \in (-19,-22)$, 
which concentrate most of the mass, 
correspond to a narrow interval of the scaling exponent, namely, $\g \in (1.75,1.83)$, 
and hence a narrow interval of $D_2$. 
The concentration of mass in a small range of fractal dimensions  
is indeed expected in a multifractal \citep{Harte,Falcon}. The consequences of the 
phenomenon of mass concentration for a general statistical analysis are 
considered in the next section.

\section{Fractional statistical moments}
\label{frac}

The analysis in the preceding section is based on the order two statistical moment
$\mu_2(v)$ and, specifically, on the variance $\mu_2(v)-1$. 
However, the variance is a sufficient statistic only for the normal distribution. 
Some conclusions about the importance of other moments can be drawn in general, 
without appeal to scaling arguments. 
 
It has been remarked earlier that a positive random variable with a rms dispersion 
equal to its mean value is not nearly normal. 
From the general statistical moment inequality 
$\mu_k \geq \mu_2^{k-1},\; k \in \mathbb{N}$ \citep[p.\ 55]{ineq},
we deduce
a lower bound to the {\em skewness}:
$$
\frac{\langle {\d M_v}^3\rangle}{\langle M_v \rangle^3} \geq (\mu_2 - 1)(\mu_2 -2).
$$  
Hence, 
$\mu_2 \gg 1$ implies that 
the probability distribution function is very skewed. 
The lognormal distribution is an example 
that goes from nearly normal for $\mu_2 \gtrsim 1$ to very skewed for $\mu_2 \gg 1$. 
Furthermore, this distribution is {\em heavy-tailed}, namely, not exponentially bounded, 
so its moment generating function is ill-defined
\citep{lognormal}. For such distributions, high-order moments are 
not meaningful. 
Actually, heavy-tailed probability distributions may not 
have high-order integral moments, but they can be analyzed using 
{\em  fractional} low-order moments \citep{Shao}.
This type of analysis connects with the analysis of multifractals, which are
mass distributions with very skewed probability functions and are related 
to the lognormal model \citep{Jones-hetero,I-ApJ}.

Indeed, a complete multifractal analysis 
requires the full set of fractional $q$-moments.  
In the standard method of 
coarse multifractal analysis \citep{Harte,Falcon},  
a region that contains the mass distribution is 
coarse-grained with a partition in cells of linear size $l$. Using this partition, 
fractional statistical moments are defined as
\begin{equation}
\M_q(l) = \sum_i \left(\frac{m_i}{M}\right)^{q},\;\;q \in \mathbb{R}, 
\label{M_q}
\end{equation}
where the index $i$ runs over the set of non-empty cells, 
$m_i$ is the mass in the cell $i$, and $M= \sum_i m_i$ is the total mass. Then,  
multifractal behavior is given, for 
$l \ra 0$, by
\begin{equation}
\M_q(l) \sim l^{\tau_q},
\label{tau_q}
\end{equation}
and the {\em R\'enyi dimension} spectrum is given by 
\begin{equation}
D_q = \frac{\tau_q}{q-1}
\label{D_q}
\end{equation}
(in particular, $D_2 = \tau_2$). 

The fractional moments of the mass distribution, namely, 
$$\mu_q(v) = \frac{\langle {\rr_v}^q\rangle}{\langle \rr_v \rangle^q}
= \frac{\langle {M_v}^q\rangle}{\langle M_v \rangle^q}\,,$$
are related to $\M_q$ by 
\begin{equation} 
\mu_q(v)= 
\frac{\M_q(v)}{v^{q-1}}\,,
\label{M-mu_q}
\end{equation} 
where $v$ is the cell volume 
\citep[see, e.g.,][]{galax2}.
We restrict ourselves to moments with $q>0$, which are more interesting and easier 
to calculate from data. 
From Eqs.~(\ref{tau_q}), (\ref{D_q}) and (\ref{M-mu_q}), 
the scaling exponent of $\mu_q(v)$ is 
\begin{equation} 
\tau_q/3-(q-1)=(q-1)(D_q-3)/3.
\label{mu-D_q}
\end{equation} 
Let us remark that this exponent is negative for $q>1$, so that $\mu_q(v)$ grows 
as $v$ shrinks, and the probability function becomes more skewed.
%

Conversely, $\mu_q(v)\ra 1$ as $v \ra \infty$. 
Naturally, 
{\em cumulants}, which express the departure from Gaussianity,  
are useful in this limit and are standard in cosmology; e.g., the skewness 
and the {\em excess kurtosis} 
\citep{Peebles}.
However, cumulants are not useful to study very skewed probability distributions,  
such as the mass distribution in the strong 
clustering regime. 
If we want to replace $\mu_q$ with some moment that vanishes at homogeneity, 
we may 
consider the {\em central absolute moments}, defined as
$$\frac{\langle {|\d M_v|}^q\rangle}{\langle M_v \rangle^q},\;\;q \in \mathbb{R},$$
or consider just $\mu_q(v)-1$, both of which generalize the variance to $q \neq 2$ 
(but are not equal).

Actually, 
there is more information about the multifractal properties 
of a mass distribution in its $q$-moments with $q$ non-integer and close to one than in 
its $q$-moments with integer $q>1$. This is because the mass is concentrated 
on the singularities that fulfill $D_1=\a_1=f(\a_1)$, where $\a_1$ is the local
dimension corresponding to $q=1$ \citep{Harte,Falcon}. However, $\mu_1=1$ (at any scale), 
and $\mu_q$ is close to one for $q$ close to one. 
Therefore, 
we have to take $q$ not too close to one to have a measurable change. 
Let us see what happens in one example. 

Taking $q=1.5$ and employing the sample with absolute magnitudes 
$M_\mathrm{r} \in (-21,-22)$, for example, we find that 
both $\mu_{1.5}(v)-1$ and the corresponding central absolute moment
do have 
scaling ranges that are almost as large as possible, with exponents $-0.54$ and $-0.41$, 
respectively. 
However, on closer inspection, the former scaling splits into one part 
in the strong clustering regime, with exponent $-0.51$, and another 
in the quasi-homogeneous regime, with exponent $-0.57$. Both these exponents can be explained. 
The first value, using Eq.~(\ref{mu-D_q}) for the exponent of 
$\mu_{1.5}(v) \approx \mu_{1.5}(v)-1$,  gives $D_{1.5}=-0.06$, that is to say, 
it corresponds to the vanishing dimension of a set of isolated points. 
The exponent $-0.57$ should give, according to Eq.~(\ref{mu-D_q}), 
$D_{1.5}=-0.42$ (now hardly compatible with a vanishing dimension); 
but it is not sensible to calculate a fractal dimension only in the quasi-homogeneous regime. 
Nevertheless, we can understand the value of the exponent as follows.  

In the quasi-homogeneous regime, any mass distribution approaches normality. 
This is realized, in particular, by the lognomal model, whose moments are easily calculated \citep{lognormal}, giving
\begin{equation} 
\mu_{q} = e^{q(q-1)\s^2/2}.
\label{lnorm}
\end{equation} 
For small $\s^2$, $\mu_{q} \gtrsim 1$ and we obtain the ratio
\begin{equation}
\frac{\mu_{q} -1}{\mu_{2} -1} \approx \frac{q\,(q-1)}{2}\,.
\label{lnorm-ratio}
\end{equation} 
Indeed, the exponent $-0.57$ of $\mu_{1.5}(v)-1$ is close to $-0.61$, which is 
the exponent of $\mu_{2}(v)-1$ (equal to $-\g/3$, with $\g$ from Table~\ref{VLSresults}).
Moreover, if we compute the ratio $(\mu_{1.5}(v)-1)/(\mu_{2}(v)-1)$  
along the quasi-homogeneous interval, then we find that it goes from $0.33$ to $0.38$, in progressively better accord with Eq.~(\ref{lnorm-ratio}),
which gives $0.375$.

Regarding the scaling of 
$$\langle |\d M_v|^{1.5}\rangle/\langle M_v \rangle^{1.5},$$ 
with exponent $-0.41$, 
we have no interpretation, but it is evident that it 
cannot correspond to a sensible value of $D_{1.5}$: let us use expression 
(\ref{mu-D_q}), recalling that $D_q$ is a non-increasing function of $q$ and $D_2 > 1$. 


Since $\mu_{1.5}(v)-1$ or the central absolute moment have
no sensible scaling in the quasi-homogeneous regime, 
we may try to find the scaling of just $\mu_{1.5}(v)$ in 
the strong clustering regime. 
Here and henceforth we employ the sample with absolute magnitudes 
$M_\mathrm{r} \in (-21,-22)$.
A fit in 
the moderately strong clustering regime yields the exponent $0.19 \pm 0.04$ and hence 
$D_{1.5} = 1.8 \pm  0.2$. This value is not very precise but is in accord with the 
also quite imprecise value 
deduced from 
the multifractal spectrum $f(\a)$ 
found by \citet{I-SDSS}. 

The dimension of the mass concentrate, $D_1$, cannot be calculated as the 
$D_q$ with $q \neq 1$, 
because $\M_1=\mu_1=1$. However, the limit $q \ra 1$ is easily taken in Eq.~(\ref{D_q}), 
using Eqs.~(\ref{M_q}) and (\ref{tau_q}), and it yields the so-called 
{\em entropy dimension}
\begin{equation} 
D_1 = \lim_{l \ra 0}\frac{\sum_i\frac{m_i}{M} \log \frac{m_i}{M}}{\log l}
\end{equation} 
\citep{Harte}.
With a linear fit of the numerator of this formula versus $\log v$, 
again in the moderately strong clustering regime,
we obtain $D_{1} = 1.9 \pm  0.2$. It is somewhat smaller than the value found 
by \citet{I-SDSS}, but it is compatible with it. 

Statistical moments with integer $q >2$ 
give us little information. 
Of course, low-order cumulants are useful in the quasi-homogeneous regime.
Moreover, we observe that the skewness 
has better scaling behavior than, for example, 
$\mu_{3}(v)-1$ or the corresponding central absolute moment. 
The scaling of higher-order cumulants might play a role.
Nevertheless, 
the relevant information about the fractal regime is provided by
fractional low-order moments.

Although the above quoted results refer to the sample with 
$M_\mathrm{r} \in (-21,-22)$, we have conducted an exploration of general scaling 
properties, for other samples and values of $q$, finding that the slight
rise of scaling exponent with luminosity already seen for $q=2$ in Sect~\ref{cell_var} 
holds in general. 
This general rule confirms the expected multifractal behavior, namely, 
the smaller values of $D_q$ for more luminous galaxy subsamples 
(see Sect~\ref{VL}).
At any rate, the change of dimension is quite small for the subsamples with 
$M_\mathrm{r} \in (-19,-22)$, which concentrate most of the stellar mass. 
This quasi-uniformity of scaling properties, which reflects the phenomenon of concentration 
of mass in multifractal geometry, justifies us to consider all the galaxies at once 
for some purposes and, in particular, to study the angular projection of the 
full apparent-magnitude sample, without concern about mixing 
galaxies in a broad range of luminosities.

\section{Fractal projections}
\label{Fp}

The study of fractal projections has tradition in mathematics \citep{Falcon,FFJ}. 
In cosmology,
the properties of the angular projection of a fractal set have been considered in regard to 
the possibility of a fractal universe with no transition to homogeneity  
\citep{Mandel,Cole-Pietro,Durrer}. Of course, the study of the angular projection of the 
distribution of galaxies, that is to say, of angular galaxy surveys, is old and predates 
the study of redshift surveys \citep{Tot-Ki,Groth}. 
In fact, \citet{Peebles} uses basic properties of 
the angular distribution of galaxies as an argument against a fractal universe with 
no transition to homogeneity. The argument  
is also applicable to the stellar mass distribution.

To study the angular distribution of stellar mass, 
we have to consider 
the generalization of the theory of 
projection of fractal sets to the theory of projection of fractal {\em measures}, 
which has been developed more recently \citep{FFJ}. This is a necessary step because 
the large-scale mass distribution has to be treated 
not as a set but as a measure,   
the measure being the mass. 
A useful concept is the {\em support} of a mass distribution, namely, 
the smallest {\em closed set} that contains all the mass. 
The large-scale mass distribution appears to be a multifractal mass distribution of 
{\em non-lacunar} type, that is to say, with support in the full space \citep{I-SDSS}. 
This property amounts to the absence of {\em totally empty} cosmic voids.
These concepts are explained in detail by \citet{galax2}. 

Regarding lacunarity, let us recall that 
\citet{Durrer} argued 
that the angular projection of 
a fractal set can have a vanishing lacunarity, 
by noting that the projection of a fractal with dimension 2 or higher 
is non-fractal  
and appealing to galaxy properties (apparent sizes and opacities). 
If the three-dimensional stellar mass distribution 
is already non-lacunar, then the issue is no longer meaningful.
Furthermore, the presence of a small lacunarity would also give rise to 
a non-lacunar projection, 
as a consequence of the almost certain fact that the fractal dimension of the support of the 
mass distribution in space is larger than two \citep{I-SDSS}.

The study of multifractal projections 
is not reduced to the behaviour of lacunarity. 
The main question is how to characterize the behavior of dimensions under a projection. 
This question 
has been partially answered by \citet{Hunt-K}, 
in terms of the R\'enyi dimension spectrum $D_q$. 
The answer is the natural generalization 
of the standard and intuitive result obtained for fractal sets \citep{Falcon,FFJ}:   
the dimension of the projection of a fractal set 
onto a plane (or a smooth surface), in particular, equals the dimension of 
the set if it is smaller than two and it is two otherwise 
(this statement has to be qualified for non-random fractals, 
which can have special projections along some directions).
This result is still valid for the dimension spectrum $D_q$ of a mass distribution, with 
the restriction that $1 <q \leq 2$ \citep{Hunt-K}. The restriction is due to 
technical reasons and may not apply to every type of mass distribution. Anyway, we are 
mostly interested in that interval and, especially, in the correlation dimension $D_2$.
Since we expect that $D_2$ is in the range 1--1.5 for the stellar mass distribution
(Sects.~\ref{z} and \ref{cell_var}), it has to be preserved under angular projections.

Here, we should notice that some authors find that $D_2 \geq 2$ \citep[table 1]{Jones-RMP}.
\Red{However, large values of $D_2$ often come from power-law fits of $\xi(r) + 1$ 
on scale ranges that extend too far  
and include a quasi-homogeneous range, thus being  
biased towards $D_2=3$, as explained in Sect.~\ref{z}.}
Moreover, all those values \Red{of $D_2$} do not refer to the stellar mass distribution but
to the galaxy number distribution.

To apply 
coarse multifractal analysis to an angular projection, we partition 
the projected region in cells of 
solid angle $\O$ and replace in (\ref{tau_q}) the length $l$ by $\O^{1/2}$. 
Now, the formula that is analogous to Eq.~(\ref{M-mu_q}) and gives the moment 
$\mu_q(\Omega)$ is
\begin{equation} 
\mu_q(\Omega)= \frac{\langle {\rr_\Omega}^q\rangle}{\langle \rr_\Omega \rangle^q}
= \frac{\M_q(\Omega)}{\Omega^{q-1}}
\label{Mmuq}
\end{equation} 
(it is assumed that $q>0$).
The expected scaling exponent of $\mu_q(\Omega)$ is, in analogy with Eq.~(\ref{mu-D_q}),
\begin{equation} 
\tau_q/2-(q-1)=(q-1)(D_q/2-1).
\label{Dq}
\end{equation} 
In particular, $\mu_2(\Omega)$ is expected to scale with exponent $D_2/2-1$.
Naturally, the exponent $(q-1)(D_q/2-1)$ is only valid provided that $D_q < 2$, 
because 2 is the maximal fractal dimension of a two-dimensional projection. 


Let us now recall and generalize some standard results about the 
angular two-point correlation function, in the light of the theory of 
fractal projections.

\subsection{The angular correlation of galaxies}
\label{angular}

The reduced two-point correlation function of the angular positions in 
a flux-limited sample is denoted by $w(\th_{12})$, 
where $\th_{12}$ is the angular distance between two points. This function 
can be expressed as an integral of the two-point correlation function of 
ordinary positions $\xi(r_{12})$ over the radial coordinates $r_1$ and $r_2$.
The integral can be simplified \Red{if $\xi(r)$ is a power law}, 
Eq.~(\ref{xi}). \Red{Then,} in the 
small-angle approximation, $\th \ll 1$ 
(in radians), the integral gives
\begin{equation}
w(\th) = K\,\th^{1-\g} \left(\frac{r_0}{d_*}\right)^{\g} 
\int_{-\infty}^\infty \frac{dx}{(1+x^2)^{\g/2}}\,,
\label{w}
\end{equation}
where $K$ is a nondimensional constant that depends on $\g$ and the 
radial selection function, and $d_*$ is the characteristic sample depth
\citep{Peebles}. 
\Red{The integration variable is $x=(r_{2}-r_{1})/(r_1\th)$.} 
The integral over $x$ 
is left unevaluated \Red{to show that} 
the present approximation fails if the integral is divergent, namely, if $\g \leq 1$. 
In this case, the projected correlation function is dominated by 
pairs of points at large relative radial distances. 

Of course, all the above is as applicable to the stellar mass 
distribution as to the galaxy number distribution. 
To the scaling exponent $\g$ corresponds the fractal dimension $D_2= 3 -\g$, 
so the cases  $\g >1$ or $\g \leq 1$ correspond, respectively, 
to $D_2 < 2$ or $D_2 \geq 2$, the latter being the case of non-fractal projection. 
If $\g >1$, then the projected correlation function is dominated by 
pairs of points at small relative radial distances, and 
the three-dimensional fractal structure 
is preserved in the projection.

It is also relevant 
that $K$ and the integral in Eq.~(\ref{w}), 
for $\g >1$ and not too close to one, are factors of the order of unity, so the magnitude 
of $w(\th)$ is ruled by the quotient $r_0/d_*$ \citep{Peebles}. If 
the characteristic sample depth is much larger than $r_0$,
as is normal in deep surveys, the angular correlations are small, except at very small 
angles. This means that the projected fractal structure is only observable 
at very small $\th$, where $w(\th) \gg 1$ and the mass fluctuations are large. 
%
%

Notice that the fractal 
structure appears blurred and 
the cosmic web features are hardly perceptible in angular images of the galaxy distribution; 
for example, in the image of the Lick survey 
\citep[p.\ 41]{Peebles}. Images like this one show that $w \ll 1$ and 
are proof of large scale homogeneity, namely, of a small ratio $r_0/d_*$, but they 
\Red{do not reveal the web features that are so} conspicuous 
in slices of three dimensional redshift surveys.
In the angular projection of an ideal mathematical fractal, 
the relevant features must always appear on very small angles, 
but the galaxies have discrete nature. 
Although \Red{$x$} is integrated down to zero in Eq.~(\ref{w}), 
correlation functions are actually calculated
as sums over pairs of points in a finite set. In particular, there is some pair
at a minimal distance, while the density of projected 
far points on the angular location of that pair keeps growing as $d_*$ grows. 
It is evident that uncorrelated far points must blur the small 
scale features and obliterate them at some stage.
In fact, the projection of uncorrelated pairs of points adds a sort of 
Poisson distributed component, thus making the average angular density $\rr_\Omega$ grow 
without altering its fluctuations. Therefore, the variance 
$(\d \rr_\Omega/\rr_\Omega)^2$ gets uniformly depressed, 
with no change of the scaling behavior (if it exists).

\citet[p.\ 220]{Peebles} shows log-log plots of $w(\th)$ for the Zwicky, Lick and
Jagellonian catalogs. 
In fact, only the Zwicky catalog, with limiting apparent magnitude 
$m = 15$, has a range of small $\th$ with $w(\th) > 1$. However, the absolute value 
of the slope grows in that range and tends to two, 
the value that corresponds to 
$D_2=0$, that is to say,
to a distribution of {\em isolated} points. Something similar happens in the plot
of $w(\th)$ 
for successive slices of the APM catalog with $\D m = 0.5$ \citep[p.\ 221]{Peebles}.
Analyses of modern and hence deeper catalogs seldom show any $w > 1$. For example, 
\citet{Wang-etal} only show 
a very small range of $\th$ with $w(\th) > 1$.  

In conclusion, any strong angular correlations 
are likely to be buried in a 
homogeneous background, as indeed happens in our case (next section).

\section{Angular analysis of the stellar mass distribution}
\label{angSDSS}

We employ the \Red{same} set of $529\hspace{1pt}655$ galaxies as in Sect.~\ref{VL}, 
in the angular rectangle defined therein. 
The lattice of cells and coarse-graining system are 
the same as in Sect.~\ref{VL} (from \citealt{I-SDSS}),   
without radial coordinate, that is to say, a $4 \times 3$ basic lattice 
and a sequence of binary subdivisions. 
To cover larger solid angles, 
we add a coarser lattice, namely, a $3 \times 2$ lattice. 
All the cells have an aspect ratio close to one, 
which is convenient.


According to Eq.~(\ref{w}), 
the average of $w(\th)$ over a small cell of solid angle $\O$ is proportional to 
$\O^{(1-\g)/2}$. Therefore, 
the expected scaling of the cell mass variance is 
\begin{equation} 
\mu_2(\Omega)-1 = \frac{\langle {\d M_\Omega}^2\rangle}{\langle M_\Omega \rangle^2}
= \left(\frac{\Omega}{\O_0}\right)^{-(\g-1)/2},
\label{O0}
\end{equation} 
\Red{where $\O_0$ is the solid angle at which the projected distribution 
approaches homogeneity and can be expressed in terms of $K,r_0/d_*$ and 
$\g$.}
If $\mu_2 \gg 1$, then Eq.~(\ref{O0})
is a particular case of the fractal scaling of $q$-moments (\ref{Mmuq}), 
with exponent (\ref{Dq}), in this case, 
$D_2/2-1=(3-\g)/2-1$. 





\begin{figure}
\centering{
\includegraphics[width=7.5cm]{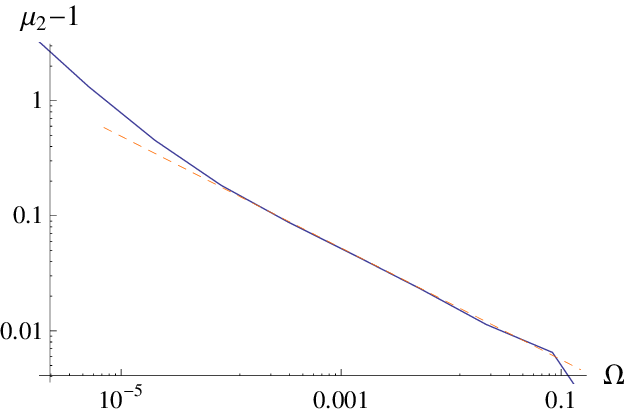}
}
\caption{
Scale-dependent variance 
of the projected stellar mass,  
with a power-law fit giving $\g=1.97$. 
}
\label{ang-corr}
\end{figure}

In Fig.~\ref{ang-corr} is the log-log plot of 
$\mu_2(\Omega)-1$ 
($\Omega$ is normalized to the total solid angle from now onwards, unless the unit 
``steradian'' appears explicitly).
The exponent of the power-law fit of $\mu_2(\O)-1$ in 
$\Omega\in [8\cdot 10^{-5},8\cdot 10^{-2}]$ turns out to be somewhat large in absolute value, 
namely, \Red{$(\g-1)/2= 0.486\pm 0.008$}, giving \Red{$\g=1.97\pm 0.02$}. 
In the fitted range, $\mu_2-1 \ll 1$, 
while $\mu_2$ grows 
for small $\O$ and $\g$ tends to three, 
due to the effect of discreteness. 

Therefore, 
in the range of $\O$ where the projected stellar mass distribution can be considered as a continuous mass distribution, it is quite uniform, with small fluctuations that are power-law correlated.
The mass fluctuations are reduced by 
the projection, 
as explained in Sect.~\ref{angular}. 
Presumably, we can have a better representation of the correlation function for small $\O$ 
by suppressing the shot noise.


\subsection{Shot noise suppression}
\label{shot}

The Poissonian fluctuations of an uncorrelated distribution of points in some volume 
are characterized by a number density variance equal to $1/N$, 
where $N$ is the mean number of points.
This term is also present when the points are correlated. 
In the case of a distribution of particles with different masses, such as galaxies, and assuming that masses are statistically independent of positions, 
the shot noise term becomes
$$\frac{\langle m^2 \rangle}{\langle m \rangle^2 N}$$
\citep[pg.~509--510]{Peebles}. 

In the coarse multifractal analysis of a sample of $N$ particles, 
when the cell size is so small that no cell contains more than one particle and therefore the sums over cells and over particles coincide, 
the calculation of the moment $\M_2$ by Eq.~(\ref{M_q}) obtains the shot noise term 
\begin{equation}
\frac{\sum m_i^2}{\left(\sum m_i\right)^2} = 
\frac{\langle m^2 \rangle}{\langle m \rangle^2 N}\,,
\label{C0}
\end{equation}
where the index $i$ of the sums runs over the set of particles.
If we start with a cell size such that no cell contains more than one particle and 
let the size grow, then, at some point, some cells contain more than one particle 
and $\M_2$ grows, because the restricted sum over these cells   
is larger than the sum over the particles in the cells 
[for a cell with two particles, $(m_1 + m_2)^2 > m_1^2 + m_2^2$]. 
Therefore, 
the term (\ref{C0}) is the minimum value of 
$\M_2$, and while $\M_2$ stays constant, $D_2=0$, 
according to Eqs.~(\ref{tau_q}) and (\ref{D_q}). At some larger scale, $\M_2$ is definitely 
growing 
and there is a crossover to a scaling with $D_2 > 0$.

\begin{figure}
\centering{
\includegraphics[width=7.5cm]{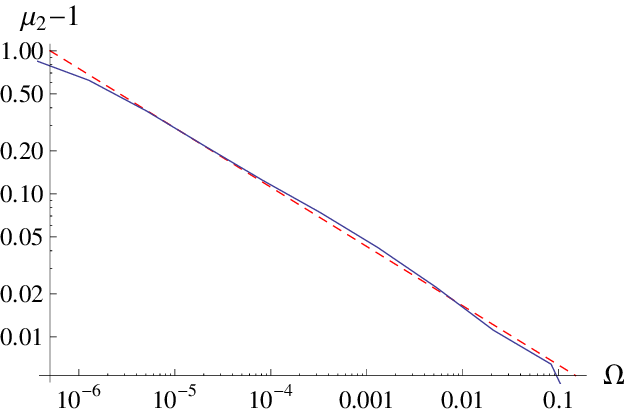}
}
\caption{Values of $\mu_2(\Omega)-1$ after shot noise suppression, 
for the distributions of 
stellar mass,  
with \Red{linear} fit (dashed line).
It gives $\g=1.83$. 
}
\label{ang-corr_1}
\end{figure}

\citet{Peebles} says about the shot noise term that
``in most applications to be discussed here, this shot noise term 
is subdominant and will be dropped'' and advises that, 
where the shot noise term is appreciable, it is to be subtracted.
We subtract the shot noise term (\ref{C0}) from $\M_2(\O)$ 
%
%
%
and see the effect in Fig.~\ref{ang-corr_1}, to be compared with 
Fig.~\ref{ang-corr}. 
It is now possible to extend 
the power-law range to $\Omega\in [1.3\cdot 10^{-6},8\cdot 10^{-2}]$,
with exponent \Red{$(\g-1)/2= 0.415 \pm 0.007$} 
(a long scaling range and a small error). 
This exponent gives $\g=1.83$, which agrees with the values found in Sect.~\ref{VL}. 
It also agrees with the results of \citet{Li-W} and of \citet{Wang-etal} (the latter for 
galaxy positions). Their procedure for calculating 
the correlation function automatically suppresses the shot noise.

\subsection{Finding the scale $r_0$}
\label{r0}

The scale $r_0$ can be found from the projected distribution 
by \Red{first} finding the angular scale of homogeneity $\Omega_0$ in Eq.~(\ref{O0}).
According to Eq.~(\ref{O0}), $\O_0$ is the solution of 
the equation $\mu_2(\Omega_0)-1=1$. 
The scale of homogeneity of the three-dimensional distribution is found 
by expressing the mass variance $\mu_2(\Omega)-1$ as the average of $w(\th)$, 
which is in turn given by Eq.~(\ref{w}). 
The equation that relates $\O_0$, $r_0/d_*$, $K$ and $\g$ follows 
from Eqs.~(\ref{w}) and (\ref{O0}) and writes 
\begin{eqnarray} 
{\O_0}^{(\g-1)/2} 
= {\O}^{(\g-1)/2} \int \frac{d\O_1 d\O_2}{\O^2}\, w(\th_{12}) \nonumber\\
= K \left(\frac{r_0}{d_*}\right)^{\g} k_\g
\int \frac{d\O_1 d\O_2}{\O^2}\left(\frac{\th_{12}}{\O^{1/2}}\right)^{1-\g}\,,
\label{O0_1}
\end{eqnarray} 
where $k_\g$ is the integral in Eq.~(\ref{w}), and 
the integral over $\O_1$ and $\O_2$ extends over a cell. 

Given $\g$, the evaluation of $k_\g$ and 
the integral over $\O_1$ and $\O_2$ are straightforward numerical computations (for this 
integral, we take into account the small size of the relevant cells to 
ignore the spherical geometry and carry out the computation in the plane). 
To calculate the constant $K$, which is an integral that contains the luminosity function, 
we can use the Schechter model \citep{Peebles}. This model has two parameters, 
\redc{namely, the characteristic galaxy luminosity $L_*$ and the slope $\a$,}
but $K$ only depends on \redc{the latter}. We take  
\redc{$\a \in (-1.2,-0.8)$, 
according to \citet{Arg} and references therein.} 

To solve for $r_0/d_*$ we have to fix $\g$ and $\O_0$. 
Taking the value of $\g$ found in Sect.~\ref{shot}, we obtain that $\O_0 = 9.53\cdot 10^{-7}$ steradians. 
With $\g=1.83$, we compute that $k_\g=3.58$ and that 
the integral over $\O_1$ and $\O_2$ is $2.33$, \redc{while $K \in (0.89,1.17)$ 
(the uncertainties in $k_\g$ and the integral are negligible, in comparison).} 
Employing Eq.~(\ref{O0_1}) \redc{(and neglecting the uncertainty in $\O_0$)}, we obtain 
\redc{$$r_0/d_*=0.0134 \pm 0.0010$$
(where the uncertainty is small enough to assume that it is normally distributed).}

To estimate the characteristic sample depth $d_*$, 
which is proportional to the square root of $L_*$, 
we take $M_*$, the absolute magnitude corresponding to $L_*$, to be 
\redc{$$M_* - 5 \log_{10}\! h = -20.6 \pm 0.2.$$
This value is based on the analysis of \citet{Arg} and 
is in the interval of magnitudes of the most representative galaxies by mass density
(Table~\ref{VLStable}).}
We obtain
\begin{eqnarray*} 
d_* &=& 10\, \mathrm{pc}\times 10^{0.2(17.77+20.6\pm 0.2-5 \log_{10}\! h)} \\
&=& (470 \pm 40)\, \mathrm{Mpc}/h.
\end{eqnarray*} 
\redc{(the uncertainty is again small enough to assume that it is normally distributed).}
Hence, 
$$r_0 = (6.3 \pm 0.7) \, \mathrm{Mpc}/h.$$ 
This is the clustering length of Eq.~(\ref{xi}), 
in contrast with the (luminosity-dependent) homogeneity scale $v_0^{1/3}$ of Sect.~\ref{VL}, 
which is tailored to the sample geometry and coarse-graining method.  
The value of $r_0$ agrees with the value of \citet{Li-W}.

\section{Summary and discussion}
\label{discuss}

The standard scaling law of galaxy clustering is 
the power-law correlation function of galaxy positions, 
with canonical exponent $\g=1.8$ and clustering length $r_0 \simeq 5\,h^{-1}$Mpc, which 
dates back over 50 years, since   
the early analyses of the galaxy-galaxy angular correlation function $w(\th)$. 
We have shown that there are two important ingredients that are missing in most  
analyses of the galaxy distribution, namely, galaxy masses and relevant statistics other 
than the variance. 
Indeed, 
more important than the distribution of galaxy positions is the distribution of stellar mass, 
which is a proxy for the distribution of baryonic matter.  
Besides, statistics appropriate to describe strong galaxy clustering are
the fractional moments, seldom employed.
We have shown that a complete study of scaling laws 
of the distribution of stellar mass reveals novelties that are worth considering. 

Our study is based on the theory of multifractal geometry. 
This theory 
assumes scaling laws for the growth of mass fluctuations on small scales, 
which define dimensions. 
The correlation dimension $D_2$, which is only one of many dimensions $D_q$, 
derives from the scaling  
the second-order statistical moment of the mass probability function.  
This moment also yields 
the scale of homogeneity. It is usually defined as the clustering length 
$r_0$, but a coarse-graining analysis obtains a volume $v_0$ (which depends somewhat on the method). 
In Sect.~\ref{z}, we have seen that $v_0^{1/3} \simeq r_0$ but that the mass fluctuations  
at the scale 
$v_0$ are still considerable. A state of 
quasi-homogeneity is reached only for length scales that are almost an order 
of magnitude larger. 

Conversely, the strong clustering regime, 
where fractal geometry applies, takes place for length scales that are almost an order 
of magnitude smaller, close to 1 Mpc.  
Unfortunately, there is not even one galaxy, on average, 
in a volume of such diameter; especially, 
if the volume is extracted from a volume-limited sample of galaxies. 
This leads to strong discreteness effects, 
which 
appear as errors in the calculation of $D_2$ and $v_0$. 
We have indeed shown that a strict definition of 
fractal scaling leads to considerable errors, namely,  
relative errors of 20\% in $D_2$ and of 30\% in $v_0^{1/3}$.  

Nevertheless, a definition of scaling that 
encompasses the quasi-homogeneous regime leads to more precise results.
The cell mass variance decreases in this regime yet 
the scaling continues,  
allowing us to calculate a precise scaling exponent $\g$.  
Although $\g$ is somewhat dependent on 
the volume-limited sample that we consider, its growth with luminosity is expected in 
a multifractal stellar mass distribution. Indeed, galaxy luminosity 
and stellar mass are strongly correlated and the stellar mass of a galaxy is correlated 
with the local dimension of the stellar mass distribution at its position. 
However, most stellar mass is concentrated in 
a restricted range of dimensions, 
namely, $D_2= 1.2$--1.3 ($\g= 1.7$--1.8), 
which is representative of the full stellar mass distribution. 
The homogeneity scale $v_0^{1/3}$ also 
grows with luminosity, being about $15\, h^{-1}$ Mpc for the set of galaxies that 
concentrates most of the mass.

It is to be remarked that the concentration of stellar mass in a range of 
galaxy luminosities that contains a small fraction of the galaxy number density 
makes the analysis of the stellar mass distribution a subject in its own right, 
quite different from the usual analysis of the 
galaxy number distribution. Nevertheless, the values of $\g$ and $r_0$
from both analyses are compatible (we recall our calculation of $r_0$ below). 
In this regard, we agree with the results of \citet{Li-W}. 

The mass variance  
is not a sufficient statistic in the fractal regime and must be complemented with 
fractional moments $\mu_q$, for $q\in\mathbb{R}$, especially, $q \gtrsim 1$.  
The study of fractional moments 
confirms the multifractal nature of the 
stellar mass distribution but shows  
no way to redefine $q$-moments (with $q \neq 2$) that extends the multifractal scaling to  
the quasi-homogeneous regime. 
In this regime, fractional moments 
simply adopt the values that correspond to a Gaussian distribution, 
as shown using the lognormal model, which interpolates between strong clustering and 
homogeneity. 
The long scaling range of the mass variance, 
which goes across the transition to homogeneity 
into the quasi-homogeneous regime, is surely a consequence of its origin 
in the {\em linear} evolution of Gaussian initial conditions.

Even the comparatively long scaling range of the mass variance
is limited to less than two orders of magnitude in length ($v^{1/3}$), as obtained  
from our volume-limited samples.   
This limitation is due to discreteness errors. 
In an attempt to 
reduce the discreteness errors, we study the angular distribution of the full flux-limited sample, which avails
the $529\hspace{1pt}967$ galaxies in our angular rectangle.  
However, in the calculation 
of $\mu_2(\O)-1$,
many pairs of galaxies in a given angular region correspond to galaxies 
at very large distances, which are uncorrelated. 
So to speak, we enhance both the signal and the noise. 
After suppressing the shot noise, 
the range of scaling of the angular stellar mass distribution grows to 
$\Omega\in [2.5\cdot 10^{-6},1.6\cdot 10^{-1}]$~sr, with exponent $\g=1.83$. 
This range of $\O$, almost five orders of magnitude, 
is equivalent to a linear range of half of that, namely, two and half orders of magnitude.  

From the angular scale of transition to homogeneity $\O_0$, 
it is possible to calculate $r_0$ 
by expressing $\O_0$ in terms of $r_0$ and 
several calculable magnitudes.  
In particular, this calculation involves 
the characteristic parameters of the galaxy luminosity function, 
which are somewhat uncertain.   
We obtain a reasonable value of $r_0$, in the range 5.8--$7.0\,h^{-1}$ Mpc.

To recapitulate and conclude, we have carried out a multifractal 
analysis of the stellar mass distribution, combining a set of 
volume-limited samples with the angular distribution.  
The methods developed here constitute an appealing alternative to other 
methods, such as methods 
that only use the galaxy positions or the two-point correlation function. 
Given that the important cosmological parameter $\s_8$ is theoretically defined 
in terms of the fluctuations of the full {\em mass} distribution, 
our methods can be relevant in the 
calculation of precision values of $\s_8$. 

	 
\section*{Acknowledgements}
I thank C.A.\ Chac\'on-Cardona for the SDSS-DR7 file. 

This research did not receive any specific grant from funding agencies in the public, commercial, or not-for-profit sectors.

\subsection*{Conflicts of interest}
The author declares that there is no conflict of interest regarding the publication of this paper.

\end{document}